\documentstyle[preprint,aps,epsfig,axodraw]{revtex}

\newcommand{\hgmn}     {{\hat{g}_{ \hat{\mu}\hat{\nu}}}}

\newcommand{\bw}     {{\beta}_W}
\newcommand{\mw}     {{\hat{m}}_W}

\newcommand{\hetamn}     {{{\eta}_{ \hat{\mu}\hat{\nu}}}}
\newcommand{\hkp}     {{\hat{\kappa}}}

\newcommand{\hf}     {{\frac{1}{2}}}
\newcommand{\hhmn}     {{\hat{h}_{ \hat{\mu}\hat{\nu}}}}

\newcommand{\lm}     {\lambda}
\newcommand{\thmnn}     {\tilde{h}_{\mu\nu}^{\vec{n}}}

\newcommand{\vn}     {{\vec{n}}}
\newcommand{\kp}     {\kappa}
\newcommand{\no}     {\nonumber}

\newcommand{\es}	{\epsilon}

\begin{document}
\draft
\preprint{
\vbox{ \hbox{SNUTP\hspace*{.2em}99-021}}
}

\title{ 
Polarization Effects on the $e^+ e^- \to W^+ W^-$ process \\
with Large Extra Dimensions }

\author{
$^{(a)}$Kang Young Lee\thanks{kylee@ctp.snu.ac.kr}, 
$^{(a,b)}$H.S. Song\thanks{hssong@physs.snu.ac.kr}, 
and  $^{(a)}$JeongHyeon Song\thanks{jhsong@ctp.snu.ac.kr}
}

\address{
$^{(a)}$ Center for Theoretical Physics, 
     Seoul National University, Seoul 151-742, Korea \\
$^{(b)}$ Department of Physics,
     Seoul National University, Seoul 151-742, Korea
}

\maketitle
\vspace{2cm}

\begin{abstract}
We study large extra dimension effects on the polarizations of the $W$ pair 
and electron beam at the $e^+ e^-\to W^+ W^-$ process. It is shown that the 
measurements of the cross section for transversely polarized $W$ pair with 
the right-handed electron beam remarkably enhance the possibilities to see 
the low scale quantum gravity effects. Higher Linear Collider bounds on 
the string scale in this model can be obtained by using the left-handed 
electron beam.
\end{abstract}
%

In spite of the extraordinary successes of the Standard Model (SM)
in explaining all the high energy experiments with large luminosity
up to now \cite{Altarelli98}, 
there have been various attempts to search for physics beyond the SM.
Even though some astrophysical and collider measurements 
have intimated that the SM is not the whole story,
the signals have been regarded as hints, not as facts.
A major portion of the motivations to extend the SM comes
from conceptual discomfort in the theoretical viewpoints,
such as the hierarchy problem:
the SM cannot provide a satisfactory answer to why the nature
allows such an enormous ratio between two fundamental mass scales,
the electroweak scale at $\sim 100$ GeV and the Planck mass scale
at $\sim 10^{19}$ GeV.
This problem has prompted the extensive studies of the supersymmetric
models \cite{Susy}, the technicolor  models \cite{Technicolor}, etc.

Recently Arkani-Hamed, Dimopoulos, and Dvali (ADD) 
have approached the hierarchy problem
in a salient way by removing one prerequisite of the problem itself:
the Planck mass is not fundamental, that is, 
there exists only one fundamental mass scale, $M_S$, in the nature \cite{ADD1}.
The observed extremely small Newton's constant $G_N$
or alternatively huge Planck mass scale can be obtained by
introducing the existence of extra compact $N$ dimensions.
Then the Planck mass is related to the $M_S$ and the size of 
extra dimensions.
For example, if the extra dimensional space is the $N$-dimensional 
torus with the same compactification radii $R$, the relation is
\begin{equation}
\label{Planck}
\kappa^2 R^N = 16 \pi (4 \pi)^{N \over 2}  \Gamma\left(
\frac{N}{2}
\right)
M_S^{-(N+2)}
\,,
\end{equation}
where $\kappa^2\equiv 16 \pi G_N$ \cite{Han}.
The excellence of Newtonian mechanics in describing the solar system excludes
the $N=1$ case where the $R$ is order of $10^{13}$ cm.
The $N=2$ case which implies mm scale extra dimensions 
is not excluded by the current macroscopic measurement of 
gravitational force \cite{macro}.
The cases of $N>2$ are also acceptable but difficult to probe through
macroscopic observations.

In order to be phenomenologically consistent with
the SM at the electroweak scale, the model
assumes that the SM fields are confined in our 4-dimensional 
brane while gravitons are freely propagating in the whole $(4+N)$-dimensional
bulk.
Even though this discrimination can be achieved by considering
our world as a topological defect of a higher dimensional theory
and by localizing the SM fields at the vortex \cite{ADD1}, 
its natural realization had been already 
discussed in string theories \cite{ADD2,Sundrum}.
Strong string coupling obtained by T-dualising
transforms the Kaluza-Klein modes of opens strings into
the winding modes of open strings of which the two ends are attached on 
a brane, which are good candidates of the SM particles.
And closed strings are still able to propagate orthogonal to the brane,
identified as gravitons.

The existence of extra dimensions reveals through the interactions
between gravitons and the SM particles on our brane.
Although the required invariance under general coordinate 
transformations in the brane and bulk
can, in principle, specify the interactions, 
the presence of non-trivial metric in the extra dimension
complicates them.
Unless gravitons take momentum compatible with or larger than $M_S$
at high energy collisions,
the spacetime region where collisions occur can be regarded as flat 
\cite{Wells}.
Then the linear approximation is valid so that
the metric in $(4+N)$-dimensions can be expanded around the Minkowski
metric as
\begin{equation}
\label{2}
\hgmn = \hetamn +\hkp \hhmn
\,,
\end{equation}
where the $\hhmn$  is the canonical graviton fields and the hatted indices
denote the $(4+N)$-dimensional spacetime.
After the compactification on a sub-manifold of extra dimensions,
our brane effectively possesses Kaluza-Klein towers of 
massive spin 2 gravitons $\thmnn$, 
massive vector particles $\tilde{A}_{\mu i}^\vn $, 
and massive spin zero particles $\tilde{\phi}_{ij}^\vn $.
The matter-graviton couplings are specified by the minimal coupling of gravity,
yielding the Feynman rules of the interactions 
between the Kaluza-Klein gravitons and the SM 
particles
\cite{Sundrum,Han,Wells}.
Following the results of Ref.\cite{Han},
we use the effective action to the leading order in $\kappa=1/M_{\rm pl}$ 
\begin{equation}
\label{3}
I=-\frac{\kappa}{2}
\sum_\vn
\int d^4 x
\left[
\tilde{h}^{\mu\nu,\vn} T_{\mu\nu} + w \tilde{\phi}^\vn T^\mu_\mu
\right]
\,,
\end{equation}
where $w=\sqrt{2/3(n+2)}$ and $T^{\mu\nu}$ is the energy-momentum tensor.
It is to be noted that new kinds of interactions due to the low scale
quantum gravity are neutral current ones.

Recently the idea of the existence of
large extra dimensions has drawn quite explosive attentions of
particle physicists. 
Above all, attractive is that the quantum gravity scale can be as low 
as TeV so to be testable.
It would be worthwhile to search for collider tests
which can confirm or exclude the validity of the model.
Various phenomenological studies have been performed to
constrain the scale of $M_S$ or the number of extra dimensions $N$
from the existing data of collider experiment \cite{Peskin,Hewett,Rizzo}, and 
in the future experiments \cite{Hewett,Rizzo,top,Agashe,Balazs,Cheung}.
Two kinds of processes at colliders are computable by using the
effective action in Eq.(\ref{3}), 
single graviton emission and virtual exchange
of gravitons.
For the first case, the presence of extra dimension 
induces the Kaluza-Klein multiplicity of phase factor,
proportional to $(\Delta E \cdot R)^N$,
where $\Delta E$ is the energy of emitting graviton.
Thus any process involving single graviton emission 
shows the large dependence on $N$ since the branching ratio is proportional to
$(\Delta E/M_S)^{N+2}$ \cite{ADD1}.
Studies of various processes
show that even in the future NLC or LHC,
the $N\geq 5$ cases are practically impossible to
distinguish from the SM background \cite{Hewett,Rizzo,top,Agashe}.
Virtual exchange of gravitons does not have 
such dependence; 
the scattering amplitudes, 
suppressed by $\kappa^2$ and mediated by Kaluza-Klein towers,
are proportional to $1/M_S^4$.
Their dependences on the $N$, according to compactification models,
may not be as critical as in the single graviton emission case 
\cite{Hewett,Rizzo}. 
Moreover in this case,
the polarizations of incoming and outgoing particles are affected by 
the fact that gravitons are of spin two while photons or $Z$ bosons,
mediating the SM neutral currents, are of spin one.
Therefore, one of the most useful and significant measurements to
probe the existence of the extra dimensions would be the measurements of
the polarizations at virtual graviton exchange processes.

In this paper we concentrate on the process $e^+ e^- \to W^+ W^-$
in the future Linear Colliders (LC) \cite{LC}.
The spin-one nature of the $W$ bosons provides 
more channels to signal new physics, and the measurements
of the $W$ polarizations are attainable 
from the decay angular distributions \cite{Wpol}.
Therefore it is desirable to derive in detail the effects 
of large extra dimensions on the $W$ and beam polarizations.

\vskip 1.5cm


\begin{center}
\begin{picture}(230,100)(0,0)
\Text(55, 5)[]{(a)}
\Text(15,100)[]{$e^-$}
\Text(15,20)[]{$e^+$}
\Text(40,60)[]{$\nu_e$}
\ArrowLine(10,90)(50,90)
\ArrowLine(50,90)(50,30)
\ArrowLine(50,30)(10,30)
\Photon(50,90)(90,90){3}{7}
\Photon(50,30)(90,30){3}{7}
\Text(95,100)[]{$W^-$}
\Text(95,20)[]{$W^+$}
\Text(170, 5)[]{(b)}
\ArrowLine(120,90)(150,60)
\ArrowLine(150,60)(120,30)
\Text(125,100)[]{$e^-$}
\Text(125,20)[]{$e^+$}
\Text(170,80)[]{$\gamma,Z,\tilde{h}_{\mu\nu}^{\vec{n}}$}
\Photon(150,60)(190,60){3}{5}
\Photon(190,60)(220,90){3}{5}
\Photon(190,60)(220,30){3}{5}
\Text(215,100)[]{$W^-$}
\Text(215,20)[]{$W^+$}
\end{picture}
\end{center}
\smallskip

{ }
\noindent
Figure~1: {\it Feynman Diagrams contributing to
the process $e^+e^-\rightarrow W^+ W^-$ including
large extra dimension effects.
}
\smallskip
\smallskip

For the process 
\begin{equation}
e^-(p_1,\kp)+e^+(p_2,\overline{\kp})\to
W^-(q_1,\lm)+W^+(q_2,\overline{\lm})
\,,
\end{equation}
there are four Feynman diagrams, one $t$-channel diagram mediated 
by the neutrino and three $s$-channel ones mediated by
the photon, $Z$ boson, and spin-2 gravitons as depicted in Fig.1.
At high energies where the electron mass is negligible,
there are 18 different helicity amplitudes 
${\mathcal M}^\pm (\lm,\overline{\lm})$,
where superscripts denote the electron helicities.
Since the CP-invariance relates some amplitudes such as
\begin{equation}
{\mathcal M}^\pm (\lm,\overline{\lm})
= {\mathcal M}^\pm (-\overline{\lm},-\lm)
\,,
\end{equation}
thus only 12 amplitudes are independent \cite{Denner}.
By using the helicity formalism in Ref.~\cite{HAmp} and
the polarization convention in Ref.~\cite{Denner},
we calculate the helicity amplitudes for the left-handed
electron beam as
\begin{eqnarray}
\label{LH}
{\mathcal M}^-_{++} &=&
s_\theta [
f_L\bw + \frac{g^2 s}{4 t}(c_\theta-\bw)
-2f_D c_\theta(1-\bw^2) ]
\,, \\ \no
{\mathcal M}^-_{+0} &=&
-\frac{\sqrt{2}(1+c_\theta)}{\mw}[
f_L\bw + \frac{g^2 s}{8 t}
(2c_\theta - 1-2\bw+\bw^2)
-f_D(2c_\theta-1)(1-\bw^2)]
\,, \\ \no
{\mathcal M}^-_{+-} &=&
2 s_\theta (1+c_\theta)[
 \frac{g^2 s}{8 t} -f_D
]
\,, \\ \no
{\mathcal M}^-_{0+} &=&
\frac{\sqrt{2}(1-c_\theta)}{\mw}[
f_L\bw + \frac{g^2 s}{8 t}
(2c_\theta+1-2\bw-\bw^2)
-f_D(2c_\theta+1)(1-\bw^2)]
\,, \\ \no
{\mathcal M}^-_{00} &=&
-\frac{s_\theta}{\mw^2}
[
f_L\bw (3-\bw^2) + \frac{g^2 s}{4 t}
(2c_\theta-3\bw+\bw^3)
-2f_D c_\theta(2-3\bw^2+\bw^4)
]
\,, \\ \no
{\mathcal M}^-_{-+} &=&
-2s_\theta (1-c_\theta)
[
\frac{g^2 s}{8 t} -f_D
]\,,
\end{eqnarray}
and for the right-handed electron beam we have
\begin{eqnarray}
\label{RH}
{\mathcal M}^+_{++} &=&
s_\theta[
f_R\bw - 2f_D c_\theta(1-\bw^2)
]
\,, \\ \no
{\mathcal M}^+_{+0} &=&
\frac{\sqrt{2}(1-c_\theta)}{\mw}[
f_R\bw
-f_D(2c_\theta+1)(1-\bw^2)
]
\,, \\ \no
{\mathcal M}^+_{+-} &=&
2 f_D s_\theta(1-c_\theta)
\,, \\ \no
{\mathcal M}^+_{0+} &=&
-\frac{\sqrt{2}(1+c_\theta)}{\mw}[
f_R\bw
-f_D(2c_\theta-1)(1-\bw^2)
]
\,, \\ \no
{\mathcal M}^+_{00} &=&
-\frac{s_\theta}{\mw^2}[
f_R\bw(3-\bw^2)-2f_D c_\theta(2-3\bw^2+\bw^4)
]
\,, \\ \no
{\mathcal M}^+_{-+} &=&
-2 f_D s_\theta(1+c_\theta)\,.
\end{eqnarray}
The CP-invariance implies
\begin{equation}
{\mathcal M}^\pm_{--}={\mathcal M}^\pm_{++}, \quad
{\mathcal M}^\pm_{-0}={\mathcal M}^\pm_{0+}, \quad
{\mathcal M}^\pm_{0-}={\mathcal M}^\pm_{+0}.
\end{equation}
In Eqs.(\ref{LH}) and (\ref{RH})
we denote $s_\theta=\sin\theta$, $c_\theta=\cos\theta$, and
\begin{eqnarray}
\es_L&=& -\hf +\sin^2\theta_W,\quad
\es_R =\sin^2\theta_W\,,
\\ \no
f_{L,R} &=& \frac{g^2 \es_{L,R}}{1-M_Z^2/s} -e^2\,,\quad
\\ \no
\mw &=& \frac{M_W}{\sqrt{s}/2} =\sqrt{1-\beta^2_W} \,.
\end{eqnarray}
The terms proportional to $f_{L,R}$ denote the contributions
from the $\gamma$- and $Z$-mediated diagrams,
and those proportional to the $g^2 s/t$ 
from the neutrino-mediated one.
It can be easily seen that the preparation of
right-handed electron beam switches off 
the $t$-channel $\nu$-mediated contributions.
These SM results are consistent with those in Ref.\cite{Denner,Agashe}.
The low scale quantum gravity effects are included in $f_D$, defined by
\begin{eqnarray}
\label{f_D}
f_D \equiv -\frac{\pi s^2}{2 M_{\rm pl}^2 } \sum_{\vn}
\frac{1}{s-m_\vn^2} 
&\simeq &
\frac{\pi s^2}{2M_S^4} \ln (
                   \frac{M_S^2}{s}
                      )
~\quad {\rm for} ~~ N=2 \\ \no & \simeq&
\frac{\pi s^2}{(N-2) M_S^4} \quad\quad {\rm for} ~~ N>2
\,.
\end{eqnarray}
It is to be noted that the two helicity amplitudes 
${\mathcal M}^+_{+-} $ and
${\mathcal M}^+_{-+} $ vanish at the tree level in the SM,
but retain extra dimension effects as sizable as
the other amplitudes.

In Fig.~2 and 3, we plot the differential cross sections for the 
$W$ pair production at $e^+ e^-$ collisions 
with respect to the $W^-$ scattering 
angle against the electron beam at $\sqrt{s}=1$ TeV
in the case of $N=2$ and $M_S=2.5$ TeV, broken down to the transverse and
longitudinal helicity components of the $W$ bosons.
The cases with $N>2$ unless $N$ is too large
shows similar behaviors.
The effects of large extra dimensions with respect to the SM background
can be enhanced by using the right-handed
electron beam and selecting the $W^+ W^-$ polarizations 
to be both transverse.
This is because the employment of the right-handed electron beam eliminates
the dominant SM contributions of
the $t$-channel $\nu$-mediated diagram in Fig.1.
And the SM background from the $s$-channel
with the right-handed electron beam is proportional to 
\begin{equation}
f_R =e^2 \left[
\frac{1}{1-M_Z^2/s}-1
\right]
\propto \frac{M_Z^2}{s}\qquad{\rm for}\quad s \gg m_Z,
\end{equation}
which $decreases$ as the beam energy becomes larger,
while the extra dimension correction proportional to $f_D$ in Eq.(\ref{f_D})
$increases$.
Furthermore, at the tree-level, 
${\mathcal M}^{+{\rm (SM)}}_{+-}={\mathcal M}^{+{\rm (SM)}}_{-+}=0$
in the SM.
The $\sigma_{_{\rm TT}}$ including large extra dimension effects
is about $\sim10^4$ times the SM background.

In practice, 
the generation of 100\% polarized electron beam is infeasible.
We consider the expected polarization of the electron beam as 90\%.
Figure 4 shows the differential cross sections 
against the $W$ scattering angle
according to the $W$ polarizations at $\sqrt{s}=1$ TeV when
$N=2$ and $M_S=2.5$ TeV.
The dominant $t$-channel SM background contaminates the unique behavior of
$d \sigma_{TT}$, however, substantial corrections to the SM background
still remain.
We do not consider the polarization of the positron beam
since it is more difficult to generate, expected presumably 
in the range of 60 $\%$ to $65\%$ \cite{LC}.

No radiative corrections are included here.
The SM radiative corrections have the effects of the same orders
of magnitude on the results with or without considering this model
since the effects of the extra dimensions 
mainly come from the interference with the SM amplitudes.

\vskip 1.0cm
\noindent
\begin{center}
\begin{tabular}{|c|cc|cc|}\hline
 &\multicolumn{2}{|c|}{~~$\sqrt{s}=0.5$ TeV 
( $\int{\mathcal L} = $50 fb$^{-1}$)~~}
&\multicolumn{2}{|c|}{~~$\sqrt{s}=1$ TeV
( $\int{\mathcal L} = $200 fb$^{-1}$)~~} 
\\ \hline
{} &~~ $N=2$~~ &~~ $N=6$~~ & ~~$N=2$~~ &~~ $N=6$~~ \\ \hline
~~$\sigma_{\rm tot}^{\rm unpol}$ ~~& 4.6 & 2.6 & 9.3 & 5.3 \\
$\sigma_{\rm TT}^{\rm unpol}$ & 4.6 & 2.6 & 9.3 & 5.4 \\
$~~\sigma_{\rm tot}^{\rm 90\%\,RH}~~$ & 3.4 & 2.0 & 6.8 & 4.1 \\
$\sigma_{\rm TT}^{\rm 90\%\,RH}$ & 3.4 & 2.0 & 7.0 & 4.1 \\
$\sigma_{\rm tot}^{\rm 90\%\,LH}$ & 5.1 & 2.9 & 10.1 & 5.8 \\
$\sigma_{\rm TT}^{\rm 90\%\,LH}$ & 5.1 & 2.9 & 10.2 & 5.8 \\
\hline
\end{tabular}
\end{center}
{Table~1}. {\it The LC bounds of $M_S$ in TeV at 95\%
confidence level according to the beam or $W$ pair polarizations.}
\vskip 0.5cm

The LC bounds on the $M_S$ are derived by the statistical errors
with the angular cut $|\cos\theta|<0.95$ at 95\% confidence level
at $\sqrt{s}=0.5$, 1.0 TeV and $N=2,6$ 
from six different observables according to
the beam and $W$ pair polarizations.
As for the beam polarization effects on the $M_S$ bounds,
the preparation of the left-handed electron beam is expected to yield 
higher bounds. 
With a given beam polarization,
the $\sigma_{_{\rm TT}}$'s are likely to give higher $M_S$ bounds.
On account of the smaller numbers of transversely polarized $W$ pair events,
these results imply that large extra dimension corrections
in the $\sigma_{_{\rm TT}}$'s are much larger than those in the
$\sigma_{\rm tot}$.

It is concluded that valuable information about the models with
large extra dimensions can be obtained 
by observing the $W$ pair and beam polarizations
at the $e^+ e^-\to W^+ W^-$ process.
In particular, 
the measurements of the cross section for transversely polarized $W$ pair
with the right-handed electron beam highly enhances the possibilities
to see the low scale quantum gravity effects.
The current inability to generate purely polarized beam at $e^+ e^-$ colliders
contaminates this feature.
Almost purely polarized beams are possible in the future $\mu^+ \mu^-$
colliders \cite{muon-collider}, 
since the muons are prepared through the pion decays
accompanied by purely chiral neutrinos.
We expect definite signal of the large extra dimension effects through 
the observations of transverse polarizations of $W$ pair
with the right-handed muon beam at the muon colliders.
It has been shown that for the LC bounds of the string scale $M_S$
the use of left-handed electron beam is preferred,
and for the probe of large extra dimension effects
the measurements of the cross section for transversely polarized $W$ pair are.

\acknowledgments

We would like to appreciate valuable discussions with S.Y. Choi
and T. Lee.
The work was supported in part by the Korea Science 
and Engineering Foundation (KOSEF) through the Center for Theoretical Physics,
Seoul National University.
HSS would like to acknowledge the financial support of the
Korea Research Foundation through the 97 Sughak Program and
98--015-D00054.

\addtocounter{figure}{1}

\vskip 4cm

\begin{center}
\begin{figure}[htb]
\hbox to\textwidth{\hss\epsfig{file=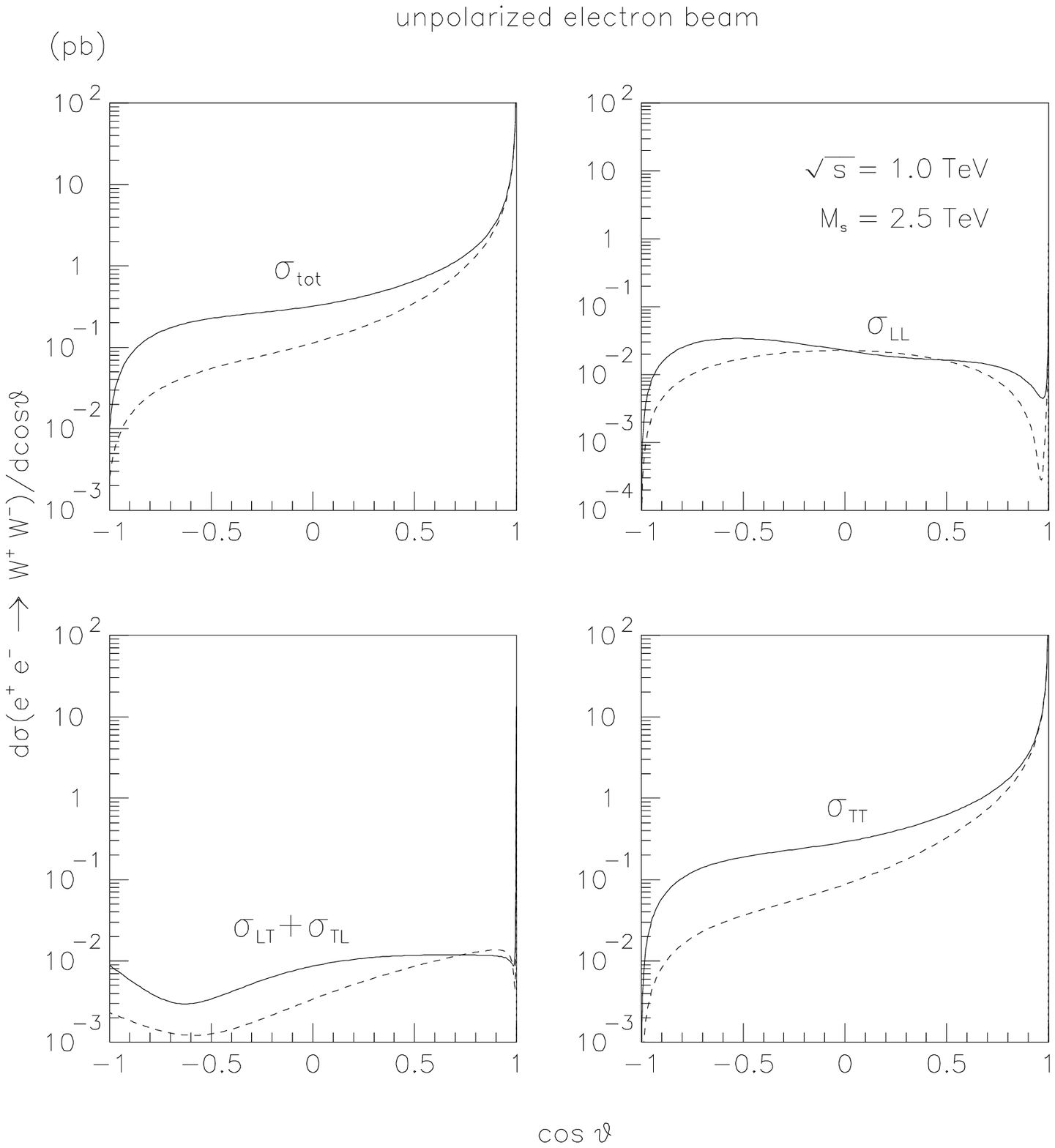,height=15cm}\hss}
\caption{\it The differential cross section with respect
		to the $W^-$ scattering angle when unpolarized beam with
		$\sqrt{s}=1$ TeV is used, broken down to the transverse 
		and longitudinal helicity components of the
		$W$ bosons.
The solid line includes large extra dimension effects when $N=2$ and
$M_S=2.5$ TeV.
The dashed line denotes the SM background.
}
\label{fig:lfig2}
\end{figure}
\end{center}

\newpage
\mbox{ }

\begin{center}
\begin{figure}[htb]
\hbox to\textwidth{\hss\epsfig{file=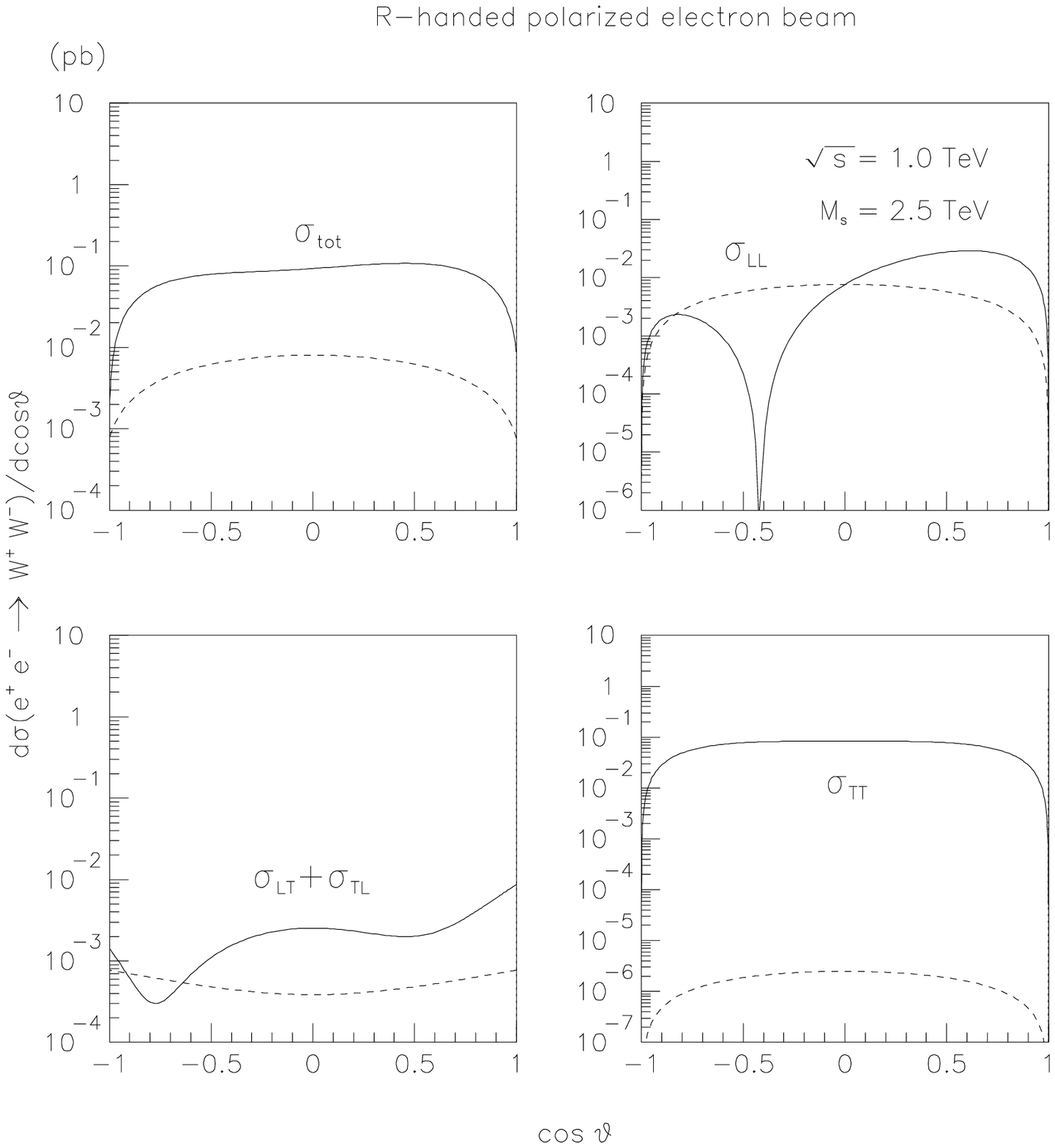,height=15cm}\hss}
\caption{\it 
The differential cross section with respect
to the $W^-$ scattering angle when the right-handed electron beam with
$\sqrt{s}=1$ TeV is used, 
broken down to the transverse and longitudinal helicity components of the
$W$ bosons.
The solid line includes large extra dimension effects when $N=2$ and 
$M_S=2.5$ TeV.
The dashed line denotes the SM background.}
\label{fig:lfig3}
\end{figure}
\end{center}

\newpage
\mbox{ }

\begin{center}
\begin{figure}[htb]
\hbox to\textwidth{\hss\epsfig{file=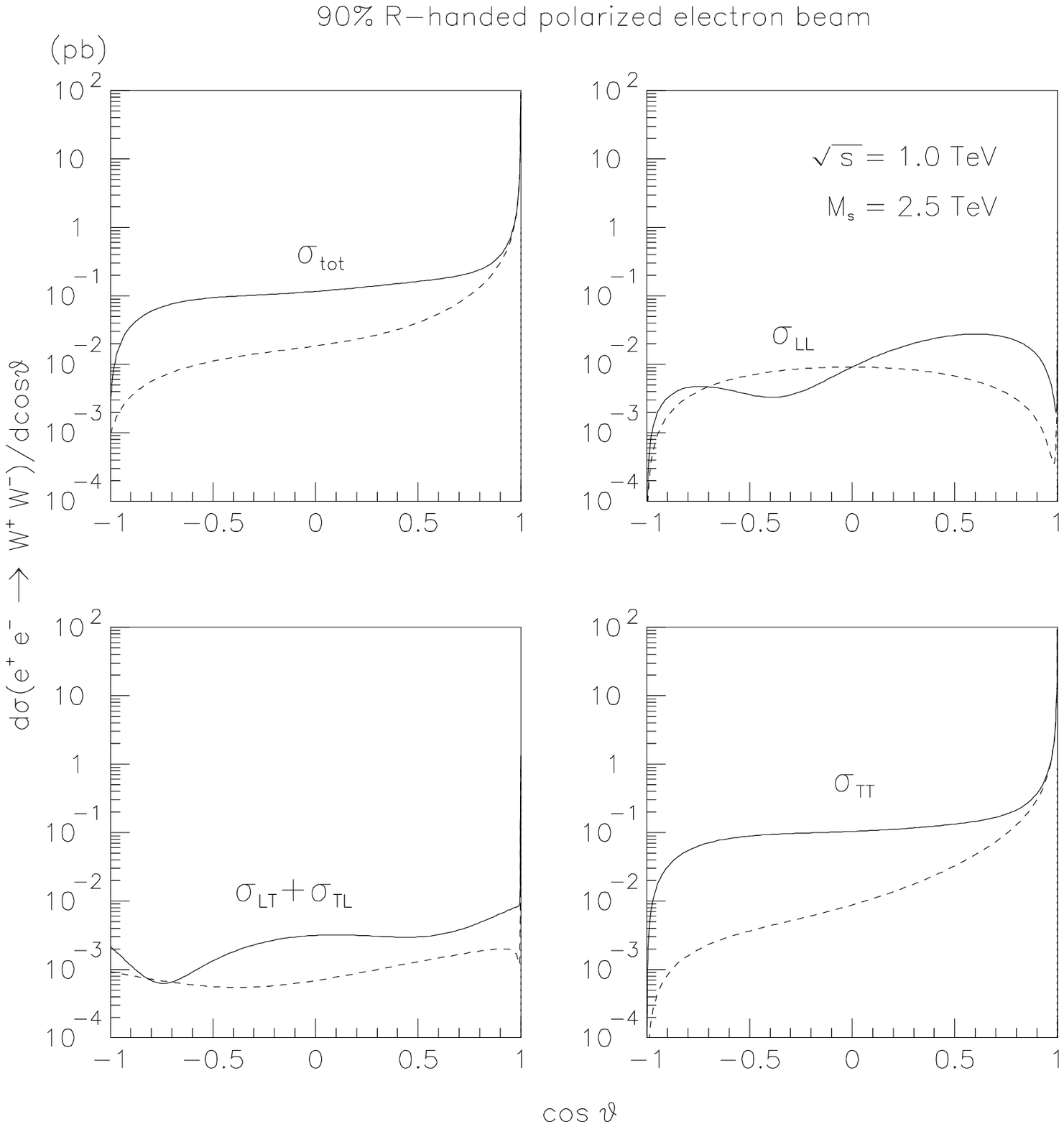,height=15cm}\hss}
\caption{\it 
The differential cross section with respect
to the $W^-$ scattering angle with the beam polarization $+90\%$ at
$\sqrt{s}=1$ TeV, 
broken down to the transverse and longitudinal helicity components of the
$W$ bosons.
The solid line includes large extra dimension effects when $N=2$ and 
$M_S=2.5$ TeV.
The dashed line denotes the SM background.
}
\label{fig:lfig4}
\end{figure}
\end{center}

\end{document}